# Room-Temperature Processing of Inorganic Perovskite Films to Enable Flexible Solar Cells

*Dianyi Liu[1], Chenchen Yang[1], Matthew Bates[1], Richard R. Lunt[1,2]\**

1. Department of Chemical Engineering and Materials Science, Michigan State University, East Lansing, Michigan 48824, USA

2. Department of Physics and Astronomy, Michigan State University, East Lansing, Michigan 48824, USA

\*Lead contact; E-mail: (rlunt@msu.edu)

ABSTRACT. Inorganic lead halide perovskite materials have attracted great attention recently due to their potential for greater thermal stability compared to hybrid organic perovskites. However, the high processing temperature to convert from the non-perovskite phase to cubic perovskite phase in many of these systems has limited their application in flexible optoelectronic devices. Here, we report a room temperature processed inorganic PSC based on CsPbI$_2$Br as the light harvesting layer. By combing this composition with key precursor solvents, we show that the inorganic perovskite film can be prepared by the vacuum-assist method under room temperature conditions in air. Unencapsulated devices achieved the power conversion efficiency up to 8.67% when measured under 1-sun irradiation. Exploiting this room temperature process, flexible inorganic PSCs based on an inorganic metal halide perovskite material is demonstrated.

Halide perovskite materials have emerged as excellent candidates for photovoltaic applications in recent years.[1-4] High device efficiency and low materials costs have given perovskite solar cells (PSCs) strong potential as a competitor for silicon solar cells.[5-7] To date, the highest reported power conversion efficiency (*PCE*) of hybrid organic-inorganic PSCs is up to 22.7%,[7, 8] which is higher than the efficiencies of polycrystalline silicon solar cells, cadmium telluride (CdTe) solar cells and copper-indium-gallium selenide (CIGS) solar cells. However, the stability issue of organic-inorganic halide perovskite materials is still a key challenge for the application of PSCs in commercial applications due to high volatility of organic components in hybrid perovskite compounds.[9-11]

In contrast, inorganic perovskite materials could have better intrinsic thermal stability.[12] Previous research suggests that pure $CsPbI_3$ perovskite can maintain the stable cubic phase over 400 °C.[12, 13] Thus, significant effort has been focused on developing of PSCs with the inorganic cesium lead halide light absorbers.[13-25] Sutton *et al.* systematically investigated the stability behavior of the cesium lead halide compounds, and reported inorganic PSCs with a *PCE* of 9.84%.[13] Chen *et al.* used the vacuum-deposition method to prepare inorganic cesium halide PSCs and achieved a device efficiency over 11%.[24] Zeng *et al.* reported a polymer-passivated cesium lead halide PSC based on the inorganic perovskite nanocrystals with the *PCE* of over 12% and an open-circuit voltage ($V_{OC}$) of over 1.3 V.[17] Wang et al very recently reported a certified efficiency of 14.67%, which is also the highest efficiency of an inorganic perovskite solar cells to date.[26] Despite the rapid research progress on improving efficiencies with inorganic PSCs, the processing of inorganic perovskite film is still challenging. Because the conversion temperature of $CsPbI_3$ from the non-perovskite phase to cubic perovskite phase occurs at over 300 °C,[12, 13] the fabrication process of $CsPbI_{3-x}Br_x$-based inorganic perovskite films generally require a thermal annealing treatment at temperatures up to 350 °C.[13, 15-20, 23, 24, 27] The high temperature thermal annealing treatment not only increases the cost of inorganic PSCs but it can prevent the application of inorganic perovskite materials on polymer-based flexible substrates.

To reduce the operation temperature of cesium lead halide perovskite films, several approaches have been examined in the past two years.[14, 28-34] For example, it was shown that doping a small amount of bromide (Br) can dramatically decrease the formation temperature of $CsPbI_3$ film.[35] Following this work, Beal *et al.* reported the low temperature processing of $CsPbBrI_2$ film as the light absorber.[14] The device was fabricated under 135 °C and obtained a *PCE* of up to 6.5%. Other efforts have introduced various additives to decrease the fabrication temperature which include hydroiodic acid, bismuth iodide, sulfobetaine zwitterions, ethylammonium iodide.[28-30, 32-34] With these additives, the cubic phase $CsPbI_{3-x}Br_x$ film could be formed under 90-150 °C. However, the thermal annealing treatment still remained an essential step for preparation of the cesium lead halide perovskite films.

Room temperature processing is not only important to simplify the fabrication procedure, but it also enables fabrication on flexible substrates.[5] To date, only two studies have reported inorganic lead halide films fabricated under room temperature that then required high temperature annealing of $TiO_2$ (450 - 500 °C) and pre-synthesized perovskite quantum dots.[25, 31]. In addition, despite many reports of flexible solar cells based on the organic-inorganic hybrid perovskite materials, flexible inorganic perovskite solar cell have not yet been reported.[5, 36-46] Here, we develop a room temperature processed inorganic PSC with $CsPbI_2Br$ as the light harvesting layer. By choosing a suitable precursor solvent, combined with the vacuum assist method, we show that inorganic perovskite films can be prepared under room temperature in air with a *PCE* up to 8.67% when measured under 1-sun irradiation. We subsequently show that this low temperature processing enables fabrication of highly flexible inorganic halide perovskite photovoltaics.

Results and Discussion

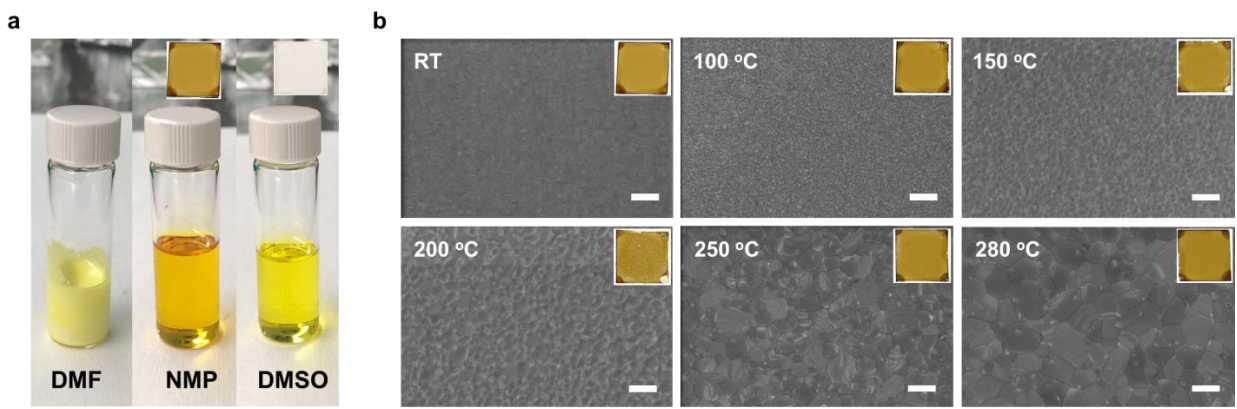

**Figure 1. Precursors Devlepment**. (a) Photograph of CsPbI$_2$Br precursor solutions prepared by various solvents. The inset shows the films prepared by the relative precursor solutions with the room temperature process. (b) SEM image of CsPbI$_2$Br films with various annealing temperatures. The inset shows the photograph of the films. The scale bar is 1 $\mu$m.

Due to the limited solubility of lead halide compounds, the precursor solvents generally chosen are N,N-dimethylformamide (DMF), dimethyl sulfoxide (DMSO) and DMF/DMSO mixtures.[2, 4, 47-50] The solubility of mixed halide cesium lead precursors are particularly limited in DMF (Figure 1a),[13] leading some researchers to utilize pure DMSO.[14, 16, 22, 29-30] However, DMSO is a Lewis-base with strong coordination capability which can result in colorless coordination complexes with lead halide compounds,[47, 51-54] and lead to difficulties in converting the lead halide perovskite precursors to the perovskite phase under room temperature (Figure 1a). Another polar aprotic solvent used to fabricate organic-inorganic hybrid perovskite solar cells is 1-methyl-2-pyrrolidone (NMP).[53, 55-58] Compared with DMF and DMSO, NMP has good solubility for cesium lead halide precursors and weak coordination affinity for lead compounds. NMP has other advantages as well including better crystallization of perovskite film and miscibility with other solvents, and has been reported as thesolvent to fabricate hybrid organic-inorganic perovskite solar cells under room temperature.[55, 57] Hence, we focus on NMP as the solvent for preparation of inorganic lead halide perovskite films with room temperature processing.

It can be seen from Figure 1a that the $CsPbI_2Br$ perovskite films can be successfully prepared by the vacuum-assist deposition process under room temperature.[48-49, 59] After the NMP solvent was extracted from the film under vacuum, the light brown $CsPbI_2Br$ perovskite film was formed. Scanning electron microscopy (SEM) images (Figure 1b) show that the $CsPbI_2Br$ film is smooth and homogenous. Due to the rapid solvent extraction the $CsPbI_2Br$ film was formed quickly with a low level of crystallization, and lead to a grain size of ~ 50 nm. To investigate the thermal stability of the room temperature processed $CsPbI_2Br$ film, we annealed the films under various temperatures in inert atmosphere. The photograph of these films clearly shows that the films maintain a brown color from room temperature to 280 °C. The color of the 100 °C film and the 150 °C film are different from the previous report by Sutton,[13] but agree with the report by Beal.[14] We infer that the film preparation method can dramatically influence the phase conversion temperature and the thermal stability of $CsPbI_2Br$ films. Accordingly, the SEM images clearly indicate that the grain size of $CsPbI_2Br$ films gradually increased from ~ 50 nm at room temperature to ~ 1 $\mu$m at 280 °C.

The as-prepared $CsPbI_2Br$ films with various annealing temperatures are also investigated by absorption spectroscopy and powder X-ray diffraction (XRD) (Figure S1). The absorption spectra show that all the films have strong absorption from 300 nm to 650 nm wavelength range (Figure S1a). After being annealed, the absorption of films presents a peak at 640 nm.

XRD patterns (Figure S1b) show that at room temperature the film has cubic perovskite peaks at $2\theta$ = 14.7, 20.9, 29.6 and 42.6° which are indexed to the (100), (110), (200) and (220) planes of $CsPbI_2Br$, respectively. After being annealed, these diffraction peaks become stronger and sharper which confirm the improved crystallization of the perovskite films. It is important to note that when the film is annealed to 150 – 250 °C, new peaks at $2\theta$ = 12.7 and 38.8° are observed. Since the new peaks are agreement with the diffraction peaks of $PbI_2$,[60] we infer that a small amount of $CsPbI_2Br$ subsequently decomposes. It is still unknown that what induced the decomposition, and a systematical investigation is needed in future work. However, these new peaks then disappear when annealing to 280 °C, which indicates the decomposition products re-react to the cubic perovskite phase.

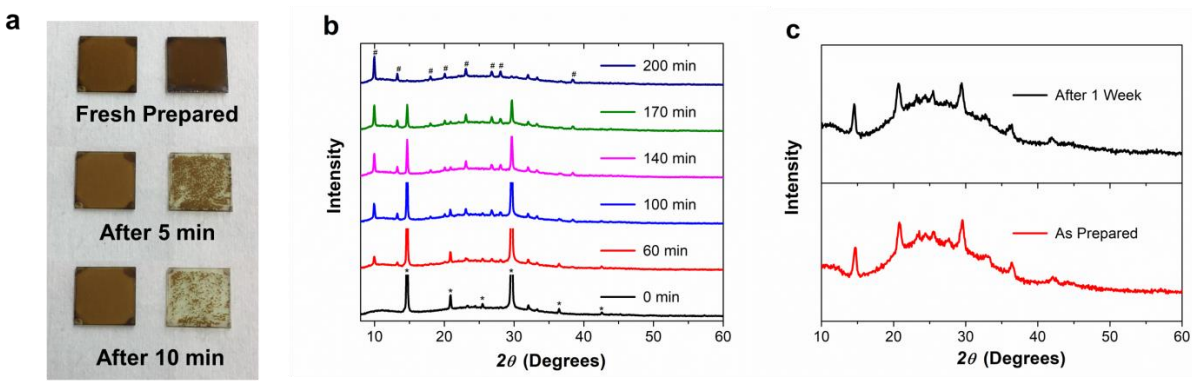

**Figure 2. Halide Perovskite Stability and Properties**. (a) Photograph of $CsPbI_2Br$ films stored under ambient air with RH = 30 ± 4%. The left row is the room-temperature film; the right row is the film with 280 °C annealing treatment. (b) XRD patterns of the film with 280 °C annealing treatment measured continuously in ambient air with RH = 22 ± 4%. (c) XRD patterns of the room-temperature film before and after storage in ambient air with RH ≤ 22 ± 4% for a week.

The humidity stability of CsPbI$_2$Br films prepared at room temperature and 280 °C was also investigated. Figure 2a is the photograph of CsPbI$_2$Br films before and after a short period (10 min) of storage under relative humidity (RH) of 30 ± 4% in open air, respectively. It is clear that the room temperature processed film can maintain the brown color. In contrast, the 280 °C film changes color in only 5 - 10 minutes, which suggests that the film undergoes a phase change or decomposition. The SEM image for the room temperature processed CsPbI$_2$Br film after storage has no obvious change (Figure S2). In comparison, the 280 °C film shows a clear change with the formation of pinholes on the film surface after storage.

To further investigate the phase change of the 280 °C film under humidity, we continously measured XRD spectra of the film under room temperature and RH = 22 ± 4% in air (Figure 2b). The initially prepared 280 °C film shows the characteristic cubic perovskite peaks. During exposure to humidity, the cubic peaks begin to fade while new diffraction peaks at $2\theta$ = 10.0, 13.3, 26.8, 28.1 and 38.4° emerge. After 200 minutes the peaks of cubic phase completely disappear. In contrast, no change is observed in the XRD spectra for the room temperature processed CsPbI$_2$Br film for over 1 week (Figure 2c). This indicates that the room temperature processed film has improved humidity stability and will therefore likely lead to improved operational lifetime as well.

Films with larger grain size are generally more compact than the films with small grain size, and the more compact film should have better stability because of the better resistance to the degradation from moisture and oxygen.[49] However, recent studies suggest that cesium lead halide perovskite films with small grain size have significantly improved stability than films with larger grain size.[28-29] The reduction in the number of pinholes and defect passivation on the surface are believed to be the main contributions to the improved stability. Our observation is consistent with these studies. Previous studies also have found that grain boundaries play an important role in the perovskite film degradation process.[61-62] Chemical residues at the grain boundaries are suggested to be one possible reason for accelerated degradation of perovskite films. Overall, the room temperature processed technique provides an effective approach to improve the humidity stability of $CsPbI_2Br$ films.

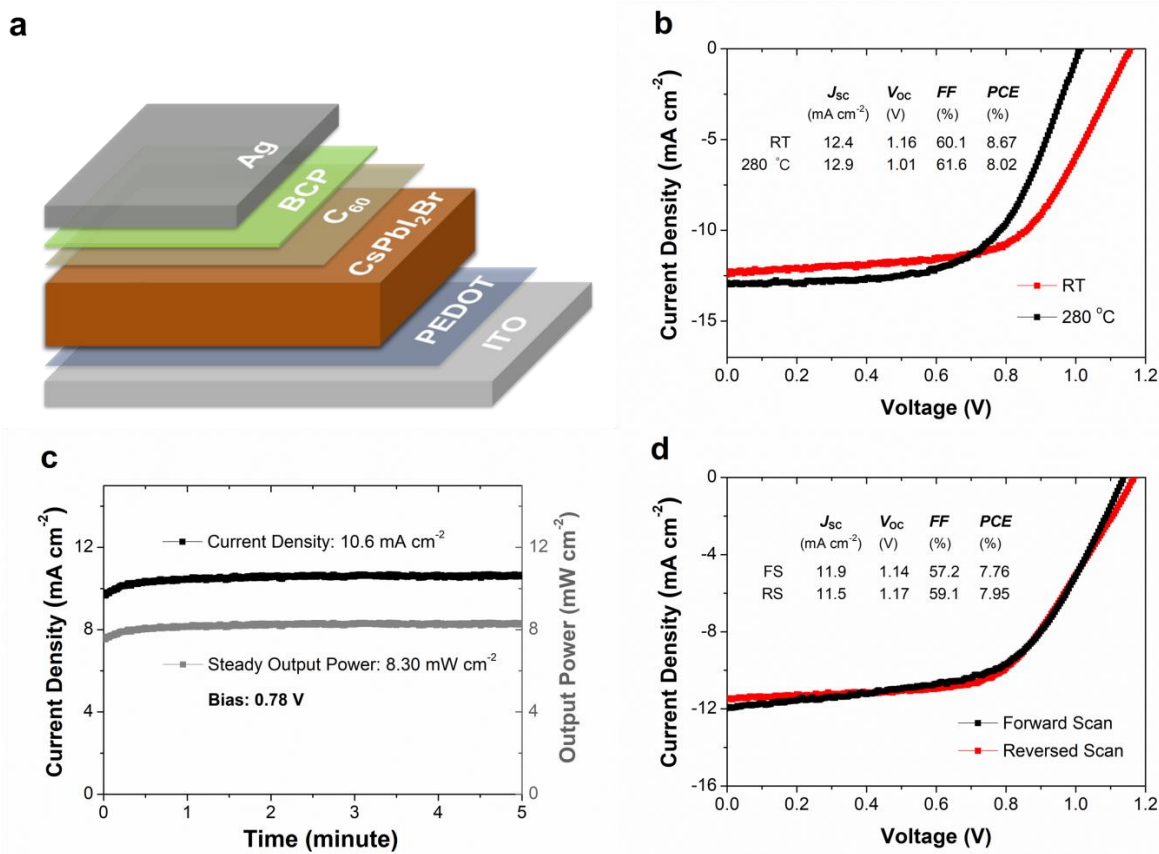

**Figure 3. Inorganic Halide Perovskite Device Performance**. (a) PSC device architecture. (b) Current−voltage (*J*−*V*) curves of perovskite devices with room-temperature $CsPbI_2Br$ film and 280 °C annealing treatment film measured under 1-sun illumination, respectively. (c) The steady current density and power output under 1-sun illumination at a bias of 0.78 V. (d) The *J*−*V* curves of room-temperature processed $CsPbI_2Br$ device measured with reverse and forward bias.

Subsequently, the room temperature processed CsPbI$_2$Br solar cells were prepared with the architechture shown in Figure 3a. A ultrathin PEDOT layer was first deposited on pre-cleaned ITO substrates. The CsPbI$_2$Br layer was then prepared on top of the PEDOT layer by the vacuum-assist method in ambient atmosphere.[48-49] A 20 nm of C$_{60}$ and a 7.5 nm of 2,9-dimethyl-4,7-diphenyl-1,10-phenanthroline (BCP) layer were then thermally evaporated onto the CsPbI$_2$Br layer, respectively, followed by the silver electrode.

The current–voltage (*J–V*) characteristics of the devices are shown in Figure 3b. Under standard AM1.5G illumination, the room temperature processed device shows a *PCE* of 8.67%, with a short circuit current ($J_{SC}$) of 12.4 mA cm$^{-2}$, a $V_{OC}$ of 1.16 V and a fill factor (*FF*) of 60.1%. The device shows an external quantum efficiency (*EQE*) spectrum (Figure S3) above 60% from 390 to 620 nm. The integrated photocurrent from the *EQE* gives a $J_{SC}$ of 12.1 mA cm$^{-2}$, which is in good agreement with the measured value from the *J-V* data. For the device annealed at 280 °C, a *PCE* of 8.02% was obtained. The lower *PCE* mainly stems from the loss of voltage. The $V_{OC}$ of the 280 °C-film device is only 1.01 V which is significantly lower than the room-temperature device. The high voltage of the room-temperature device can be attributed to fewer pinholes or shunting pathways. Figure 3c shows the steady photocurrent and power output of room-temperature device is 10.6 mA cm$^{-2}$ and 8.30 mW cm$^{-2}$ under a bias of 0.78 V, respectively. Moreover, the room temperature processed device only shows a small photocurrent hysteresis when measured under forward and reversed scan mode (Figure 3d).

Based on the room temperature processing technique, inorganic cesium lead halide perovskite device were prepared on flexible substrates (Figure 4a). An ITO/poly(ethylene terephthalate) (PET) flexible substrate was used to replace the rigid ITO/glass substrate to prepare the flexible device. Figure 4b shows the *J-V* curve of the flexible cesium lead halide perovskite solar cell.

The flexible device shows a *PCE* of 6.50% under 1-sun illumination, with a $J_{SC}$ of 12.0 mA cm$^{-2}$, a $V_{OC}$ of 1.05 V and a *FF* of 51.4%. The unencapsulated flexible device maintained an efficiency of 6.05% after 2 months storage in inert atmosphere, which is 93% of the original efficiency.

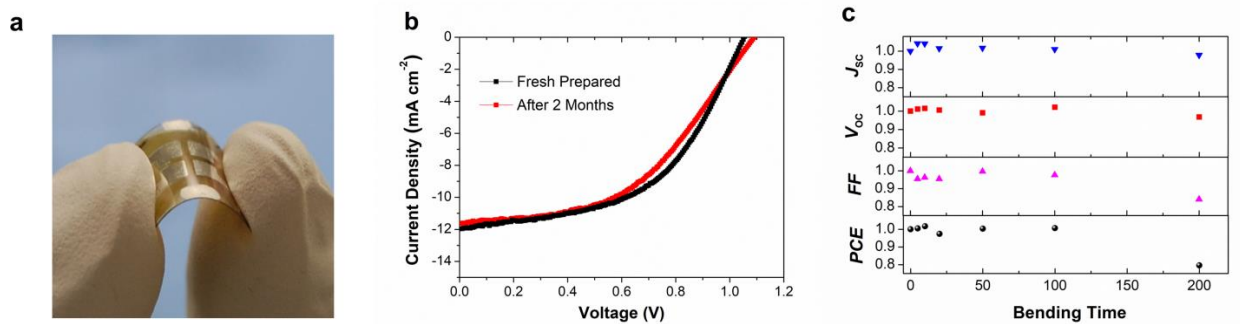

**Figure 4. All Inorganic Halide Perovskite Flexible Solar Cells**. (a) *J–V* curves of the champion flexible CsPbI$_2$Br device measured before and after 2 months storage in a glovebox without any encapsulation. (b) Normalized parameters of flexible CsPbI$_2$Br devices under various bending circles with a bending radius of 4.05 mm. (c) Photograph of the flexible CsPbI$_2$Br devices.

Bending tests were carried out to check the performance of flexible device after repeated bend cycles. After 100 bending cycles around a radius of 4.05 mm, the device only showed slight fluctuations in the efficiency (Figure 4c). However, the efficiency of the flexible device drops to ~80% after 200 bending circles, largely due to drops in the FF which likely stem from cracking of the ITO/PET flexible substrate. While the performance of flexible inorganic PSC can still be improved, these first demonstrations are encouraging for the development of low cost, flexible and stable inorganic perovskite solar cells.

**Conclusions**

In summary, we demonstrate an approach to prepare inorganic lead perovskite films with a room temperature process. The room-temperature film shows improved humidity stability over films prepared by high temperature annealing treatment. Utilizing the room temperature approach, the inorganic lead perovskite solar cells are successfully prepared on rigid substrates and flexible substrates. This work demonstrates the integration of inorganic halide perovskites into flexible solar cells and highlights the great potential of inorganic perovskite materials for a range of flexible optoelectronic devices.

**Experimental Section**

*Materials and Precursor Preparation:* 1-methyl-2-pyrrolidone (NMP, anhydrous, 99.5%, Aldrich.), dimethylformamide (DMF, anhydrous, 99.8%, Sigma Aldrich), dimethyl sulfoxide (DMSO, anhydrous, 99.9%, Sigma Aldrich), PEDOT:PSS (Clevios PVP AI 4083, Heraeus, diluted to 5% with water), $PbI_2$ (99%, Sigma Aldrich), CsBr (99.999%, Sigma Aldrich.), $C_{60}$ (99.9%, MER Corporation.) and 2,9-dimethyl-4,7-diphenyl-1,10-phenanthroline (BCP, Lumtech) were used as received.

To prepare the perovskite precursor solution, $PbI_2$: CsBr (461 mg and 213 mg, respectively) were added in NMP (1.2 ml). The solutions were stirred for 1 hour and filtered with 0.45 μm PTFE filters before use.

*Device Fabrication:* The PEDOT layer and inorganic perovskite layer were prepared under open air conditions (with a measured relative humidity ≤ 36 ± 4%). The PEDOT solution was spin-coated onto pre-cleaned ITO substrates at 6000 rpm for 20 s. The perovskite precursor was spin-coated on top of the PEDOT film at 6000 rpm for 12 s, and then moved the substrate into a homemade vacuum chamber, evacuated to ~ 10 mtorr, and left in the chamber for 3 min. The samples were then transferred into glovebox. For high-temperature films, the substrates were annealed to 260-280 °C for 1 min. The substrates were then moved into the evaporation chamber for deposition of $C_{60}$ (20 nm) and BCP (7.5 nm). Finally, an 80 nm thick silver layer was deposited by thermal evaporation at a base pressure of $3 \times 10^{-6}$ Torr through a shadow mask with a final measured device area of 4.85 mm$^2$.

*Measurement and Characterization:* The current density–voltage characteristics (*J–V* curves) were obtained using a Keithley 2420 sourcemeter under AM1.5 G solar simulation (xenon arc lamp with the spectral-mismatch factor of 0.980.) where the light intensity was measured using a NREL-calibrated Si reference cell with KG5 filter. For devices with annealing treatment, the devices were encapsulated in epoxy in glovebox before being tested due to their sensitivity to air. The room temperature films and devices, include the flexible devices, were generally tested under ambient air condition (the RH ≤ 36 ± 4%) without any protection or encapsulation. Devices were scanned at a rate of 50 mV s$^{-1}$. The *EQE* measurements were performed using a QTH lamp with a monochromator, chopper, lock-in amplifier, and calibrated Si detector to measure the intensity. XRD data was measured using CuKα (0.154 nm) emission with a Bruker D2 phaser. The *PL, UV-vis* and XRD were measured on un-encapsulated samples in ambient air (the relative humidity ≤ 22 ± 4%). A field-emission scanning electron microscopy (Carl Zeiss Auriga Dual Column FIB SEM) was used to acquire SEM images.

AUTHOR INFORMATION

**Corresponding Author**

Prof. Richard R. Lunt

*E-mail: (rlunt@msu.edu)

Department of Chemical Engineering and Materials Science, Department of Physics and Astronomy, Michigan State University, East Lansing, Michigan 48824, USA

ACKNOWLEDGMENT

The authors acknowledge financial support from the Michigan State University Strategic Partnership Grant (SPG) (D. L.) and from the U.S. Department of Energy (DOE) Office of Science, under Award # DE-SC0010472.

Supporting Information

# Room-Temperature Processing of Inorganic Perovskite Films to Enable Flexible Solar Cells

*Dianyi Liu[1], Chenchen Yang[1], Matthew Bates[1], Richard R. Lunt[1,2]\**

1. Department of Chemical Engineering and Materials Science, Michigan State University,

East Lansing, Michigan 48824, USA

2. Department of Physics and Astronomy, Michigan State University, East Lansing,

Michigan 48824, USA

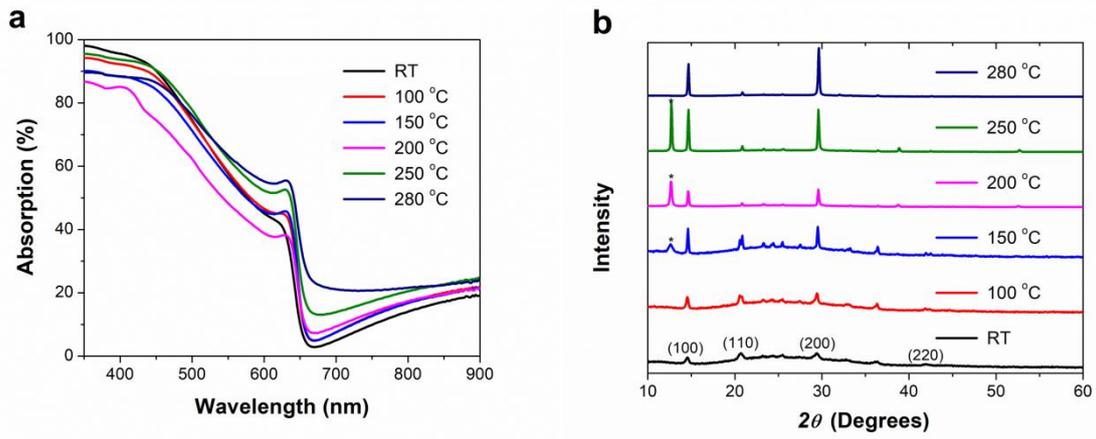

Figure S1. Absorption spectra (a) and XRD patterns (b) of CsPbI$_2$Br films with various temperature treatments.

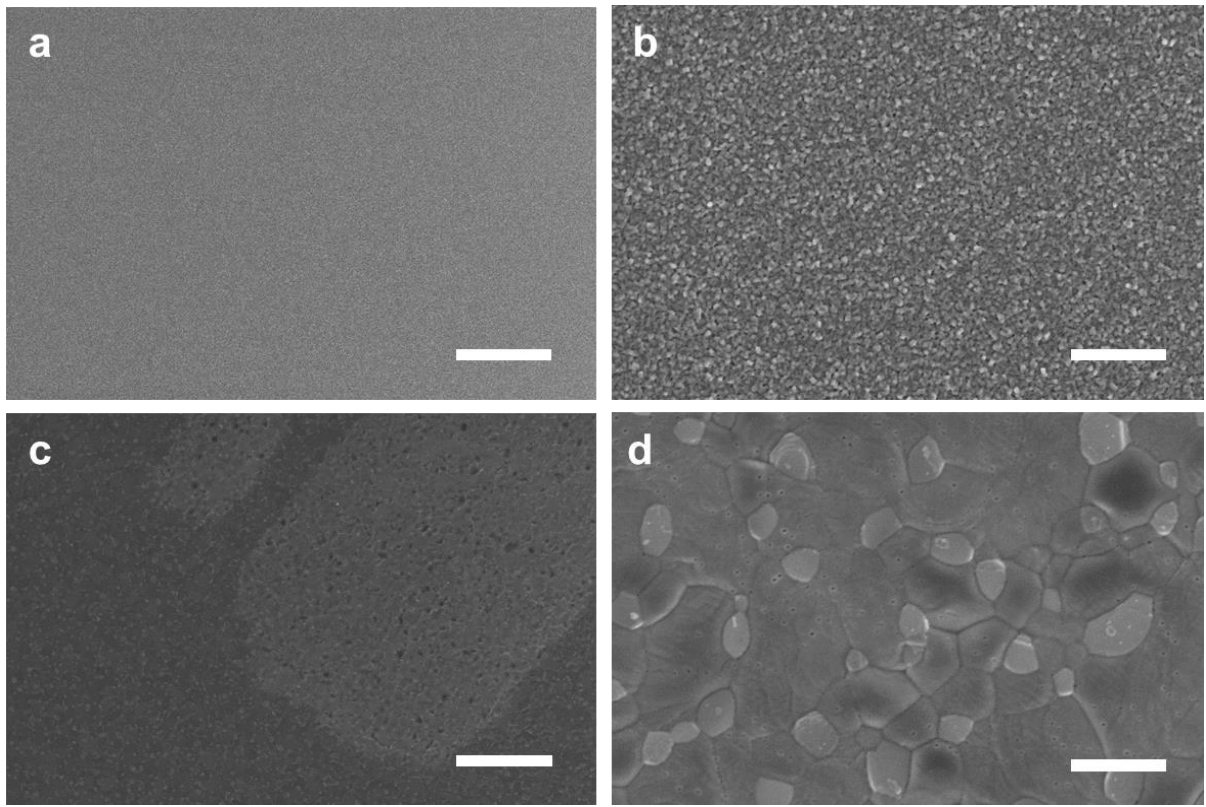

Figure S2. SEM images of CsPbI$_2$Br film stored in ambient air (RH = 30 ± 4%) for 5 minutes. (a) and (b) is the room temperature CsPbI$_2$Br film; (c) and (d) is the 280 °C annealing treatment CsPbI$_2$Br film. The scale bar is 10 μm for (a) and (c), 1 μm for (b) and (d), respectively.

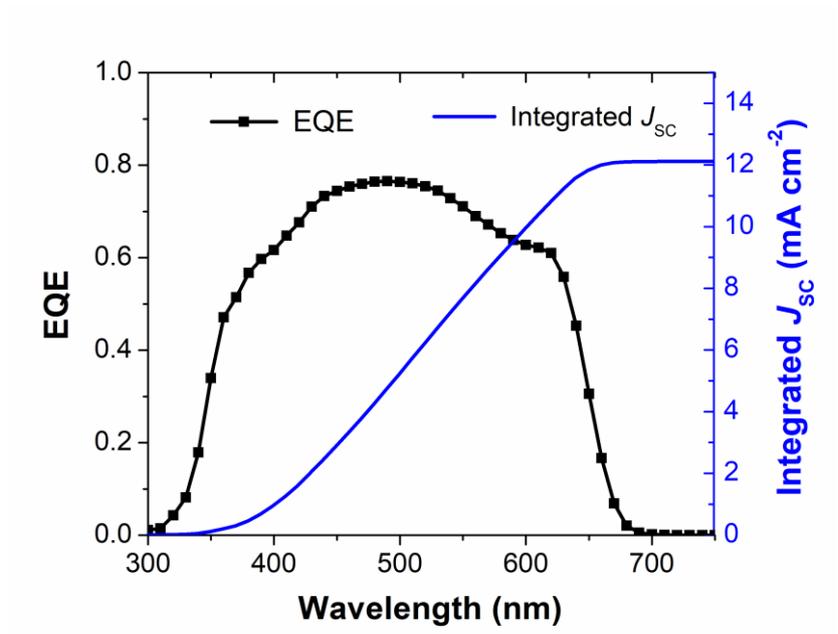

Figure S3. The *EQE* spectrum of the room-temperature CsPbI$_2$Br devices.